\pdfoutput=1

\documentclass[12pt,a4paper]{article}

\usepackage{ifthen} 
\newboolean{pdflatex}
\setboolean{pdflatex}{true} 

\newboolean{articletitles}
\setboolean{articletitles}{true} 

\newboolean{uprightparticles}
\setboolean{uprightparticles}{false} 

\newboolean{inbibliography}
\setboolean{inbibliography}{false} 

\usepackage[top=1in, bottom=1.25in, left=1in, right=1in]{geometry}

\usepackage{microtype}
\usepackage{lineno}  
\usepackage{xspace} 
\usepackage{caption} 

\usepackage{graphicx}  
\usepackage{color}
\usepackage{colortbl}
\graphicspath{{./figs/}} 

\usepackage{amsmath} 
\usepackage{amssymb}
\usepackage{amsfonts}
\usepackage{upgreek} 

\newcommand*\patchAmsMathEnvironmentForLineno[1]{%
\expandafter\let\csname old#1\expandafter\endcsname\csname #1\endcsname
\expandafter\let\csname oldend#1\expandafter\endcsname\csname
end#1\endcsname
 \renewenvironment{#1}%
   {\linenomath\csname old#1\endcsname}%
   {\csname oldend#1\endcsname\endlinenomath}%
}
\newcommand*\patchBothAmsMathEnvironmentsForLineno[1]{%
  \patchAmsMathEnvironmentForLineno{#1}%
  \patchAmsMathEnvironmentForLineno{#1*}%
}
\AtBeginDocument{%
\patchBothAmsMathEnvironmentsForLineno{equation}%
\patchBothAmsMathEnvironmentsForLineno{align}%
\patchBothAmsMathEnvironmentsForLineno{flalign}%
\patchBothAmsMathEnvironmentsForLineno{alignat}%
\patchBothAmsMathEnvironmentsForLineno{gather}%
\patchBothAmsMathEnvironmentsForLineno{multline}%
\patchBothAmsMathEnvironmentsForLineno{eqnarray}%
}

\usepackage{hyperref}    

\usepackage{xspace} 
\usepackage{upgreek}

\def\lhcb {\mbox{LHCb}\xspace}

\def\argus  {\mbox{ARGUS}\xspace}

\def\rich   {RICH\xspace}

\def\MagUp {\mbox{\em Mag\kern -0.05em Up}\xspace}

\ifthenelse{\boolean{uprightparticles}}%
{

 \def\Pmu         {\ensuremath{\upmu}\xspace}

 \def\Ppi         {\ensuremath{\uppi}\xspace}

 \def\Ppsi        {\ensuremath{\uppsi}\xspace}

 \def\PDelta      {\ensuremath{\Delta}\xspace}                 
 \def\PXi      {\ensuremath{\Xi}\xspace}                 
 \def\PLambda      {\ensuremath{\Lambda}\xspace}                 
 \def\PSigma      {\ensuremath{\Sigma}\xspace}                 
 \def\POmega      {\ensuremath{\Omega}\xspace}                 
 \def\PUpsilon      {\ensuremath{\Upsilon}\xspace}                 
 
 \def\PB      {\ensuremath{\mathrm{B}}\xspace}                 
                  
 \def\PD      {\ensuremath{\mathrm{D}}\xspace}

 \def\PJ      {\ensuremath{\mathrm{J}}\xspace}                 
 \def\PK      {\ensuremath{\mathrm{K}}\xspace}

 \def\PW      {\ensuremath{\mathrm{W}}\xspace}

 \def\PZ      {\ensuremath{\mathrm{Z}}\xspace}                 
                  
 \def\Pb      {\ensuremath{\mathrm{b}}\xspace}                 
 \def\Pc      {\ensuremath{\mathrm{c}}\xspace}                 
 \def\Pd      {\ensuremath{\mathrm{d}}\xspace}                 
 \def\Pe      {\ensuremath{\mathrm{e}}\xspace}

 \def\Pi      {\ensuremath{\mathrm{i}}\xspace}

 \def\Pp      {\ensuremath{\mathrm{p}}\xspace}

 \def\Ps      {\ensuremath{\mathrm{s}}\xspace}                 
 \def\Pt      {\ensuremath{\mathrm{t}}\xspace}

}
{

 \def\Pmu         {\ensuremath{\mu}\xspace}

 \def\Ppi         {\ensuremath{\pi}\xspace}

 \def\Ppsi        {\ensuremath{\psi}\xspace}                 
                  
 \mathchardef\PDelta="7101
 \mathchardef\PXi="7104
 \mathchardef\PLambda="7103
 \mathchardef\PSigma="7106
 \mathchardef\POmega="710A
 \mathchardef\PUpsilon="7107
                  
 \def\PB      {\ensuremath{B}\xspace}                 
                  
 \def\PD      {\ensuremath{D}\xspace}

 \def\PJ      {\ensuremath{J}\xspace}                 
 \def\PK      {\ensuremath{K}\xspace}

 \def\PW      {\ensuremath{W}\xspace}

 \def\PZ      {\ensuremath{Z}\xspace}                 
                  
 \def\Pb      {\ensuremath{b}\xspace}                 
 \def\Pc      {\ensuremath{c}\xspace}                 
 \def\Pd      {\ensuremath{d}\xspace}                 
 \def\Pe      {\ensuremath{e}\xspace}

 \def\Pi      {\ensuremath{i}\xspace}

 \def\Pp      {\ensuremath{p}\xspace}

 \def\Ps      {\ensuremath{s}\xspace}                 
 \def\Pt      {\ensuremath{t}\xspace}

}

\makeatletter
\ifcase \@ptsize \relax
  \newcommand{\miniscule}{\@setfontsize\miniscule{4}{5}}
\or
  \newcommand{\miniscule}{\@setfontsize\miniscule{5}{6}}
\or
  \newcommand{\miniscule}{\@setfontsize\miniscule{5}{6}}
\fi
\makeatother

\DeclareRobustCommand{\optbar}[1]{\shortstack{{\miniscule (\rule[.5ex]{1.25em}{.18mm})}
  \\ [-.7ex] $#1$}}

\def\epem       {{\ensuremath{\Pe^+\Pe^-}}\xspace}

\def\mumu       {{\ensuremath{\Pmu^+\Pmu^-}}\xspace}

\def\Wpm    {{\ensuremath{\PW^\pm}}\xspace}
\def\Z      {{\ensuremath{\PZ}}\xspace}

\def\dquark    {{\ensuremath{\Pd}}\xspace}
\def\dquarkbar {{\ensuremath{\overline \dquark}}\xspace}

\def\squark    {{\ensuremath{\Ps}}\xspace}
\def\squarkbar {{\ensuremath{\overline \squark}}\xspace}

\def\cquark    {{\ensuremath{\Pc}}\xspace}

\def\bquark    {{\ensuremath{\Pb}}\xspace}

\def\tquark    {{\ensuremath{\Pt}}\xspace}

\def\pion   {{\ensuremath{\Ppi}}\xspace}

\def\pip    {{\ensuremath{\pion^+}}\xspace}
\def\pim    {{\ensuremath{\pion^-}}\xspace}
\def\pipm   {{\ensuremath{\pion^\pm}}\xspace}
\def\pimp   {{\ensuremath{\pion^\mp}}\xspace}

\def\kaon    {{\ensuremath{\PK}}\xspace}
  \def\Kbar    {{\kern 0.2em\overline{\kern -0.2em \PK}{}}\xspace}

\def\KorKbar    {\kern 0.18em\optbar{\kern -0.18em K}{}\xspace}

\def\Kp      {{\ensuremath{\kaon^+}}\xspace}
\def\Km      {{\ensuremath{\kaon^-}}\xspace}
\def\Kpm     {{\ensuremath{\kaon^\pm}}\xspace}
\def\Kmp     {{\ensuremath{\kaon^\mp}}\xspace}

\def\Kstarz  {{\ensuremath{\kaon^{*0}}}\xspace}
\def\Kstarzb {{\ensuremath{\Kbar{}^{*0}}}\xspace}

  \def\Dbar    {{\kern 0.2em\overline{\kern -0.2em \PD}{}}\xspace}
\def\D       {{\ensuremath{\PD}}\xspace}

\def\DorDbar    {\kern 0.18em\optbar{\kern -0.18em D}{}\xspace}
\def\Dz      {{\ensuremath{\D^0}}\xspace}
\def\Dzb     {{\ensuremath{\Dbar{}^0}}\xspace}

\def\Dm      {{\ensuremath{\D^-}}\xspace}

\def\Dstarp  {{\ensuremath{\D^{*+}}}\xspace}

\def\Dsm     {{\ensuremath{\D^-_\squark}}\xspace}

\def\B       {{\ensuremath{\PB}}\xspace}
\def\Bbar    {{\ensuremath{\kern 0.18em\overline{\kern -0.18em \PB}{}}}\xspace}

\def\BorBbar    {\kern 0.18em\optbar{\kern -0.18em B}{}\xspace}
\def\Bz      {{\ensuremath{\B^0}}\xspace}

\def\Bu      {{\ensuremath{\B^+}}\xspace}

\def\Bp      {{\ensuremath{\Bu}}\xspace}

\def\Bpm     {{\ensuremath{\B^\pm}}\xspace}

\def\Bd      {{\ensuremath{\B^0}}\xspace}
\def\Bs      {{\ensuremath{\B^0_\squark}}\xspace}

\def\jpsi     {{\ensuremath{{\PJ\mskip -3mu/\mskip -2mu\Ppsi\mskip 2mu}}}\xspace}

  \def\Y#1S{\ensuremath{\PUpsilon{(#1S)}}\xspace}

\def\proton      {{\ensuremath{\Pp}}\xspace}

\def\Xires       {{\ensuremath{\PXi}}\xspace}

\def\Lz          {{\ensuremath{\PLambda}}\xspace}
\def\Lbar        {{\ensuremath{\kern 0.1em\overline{\kern -0.1em\PLambda}}}\xspace}
\def\LorLbar    {\kern 0.18em\optbar{\kern -0.18em \PLambda}{}\xspace}

\def\Lb      {{\ensuremath{\Lz^0_\bquark}}\xspace}

\def\Xib     {{\ensuremath{\Xires_\bquark}}\xspace}

\def\BF         {{\ensuremath{\mathcal{B}}}\xspace}

\newcommand{\decay}[2]{\ensuremath{#1\!\to #2}\xspace}         

\def\to                 {\ensuremath{\rightarrow}\xspace}

\def\Vtd  {{\ensuremath{V_{\tquark\dquark}}}\xspace}

\def\Vts  {{\ensuremath{V_{\tquark\squark}}}\xspace}

\def\AT#1     {\ensuremath{A_{\mathrm{T}}^{#1}}\xspace}           

\def\C#1      {\ensuremath{\mathcal{C}_{#1}}\xspace}                       
\def\Cp#1     {\ensuremath{\mathcal{C}_{#1}^{'}}\xspace}                    
\def\Ceff#1   {\ensuremath{\mathcal{C}_{#1}^{\mathrm{(eff)}}}\xspace}        
\def\Cpeff#1  {\ensuremath{\mathcal{C}_{#1}^{'\mathrm{(eff)}}}\xspace}       
\def\Ope#1    {\ensuremath{\mathcal{O}_{#1}}\xspace}                       
\def\Opep#1   {\ensuremath{\mathcal{O}_{#1}^{'}}\xspace}                    

\newcommand{\tev}{\ifthenelse{\boolean{inbibliography}}{\ensuremath{~T\kern -0.05em eV}\xspace}{\ensuremath{\mathrm{\,Te\kern -0.1em V}}}\xspace}
\newcommand{\gev}{\ensuremath{\mathrm{\,Ge\kern -0.1em V}}\xspace}
\newcommand{\mev}{\ensuremath{\mathrm{\,Me\kern -0.1em V}}\xspace}
\newcommand{\kev}{\ensuremath{\mathrm{\,ke\kern -0.1em V}}\xspace}
\newcommand{\ev}{\ensuremath{\mathrm{\,e\kern -0.1em V}}\xspace}
\newcommand{\gevc}{\ensuremath{{\mathrm{\,Ge\kern -0.1em V\!/}c}}\xspace}
\newcommand{\mevc}{\ensuremath{{\mathrm{\,Me\kern -0.1em V\!/}c}}\xspace}
\newcommand{\gevcc}{\ensuremath{{\mathrm{\,Ge\kern -0.1em V\!/}c^2}}\xspace}
\newcommand{\gevgevcccc}{\ensuremath{{\mathrm{\,Ge\kern -0.1em V^2\!/}c^4}}\xspace}
\newcommand{\mevcc}{\ensuremath{{\mathrm{\,Me\kern -0.1em V\!/}c^2}}\xspace}

\def\mm   {\ensuremath{\mathrm{ \,mm}}\xspace}

\def\mum  {\ensuremath{{\,\upmu\mathrm{m}}}\xspace}

\def\invfb   {\ensuremath{\mbox{\,fb}^{-1}}\xspace}

\newcommand{\chisq}{\ensuremath{\chi^2}\xspace}
\newcommand{\chisqndf}{\ensuremath{\chi^2/\mathrm{ndf}}\xspace}
\newcommand{\chisqip}{\ensuremath{\chi^2_{\text{IP}}}\xspace}

\def\gsim{{~\raise.15em\hbox{$>$}\kern-.85em
          \lower.35em\hbox{$\sim$}~}\xspace}
\def\lsim{{~\raise.15em\hbox{$<$}\kern-.85em
          \lower.35em\hbox{$\sim$}~}\xspace}

\def\sqs   {\ensuremath{\protect\sqrt{s}}\xspace}

\def\ptot       {\mbox{$p$}\xspace}
\def\pt         {\mbox{$p_{\mathrm{ T}}$}\xspace}

\def\evtgen     {\mbox{\textsc{EvtGen}}\xspace}

\def\geant      {\mbox{\textsc{Geant4}}\xspace}

\def\photos     {\mbox{\textsc{Photos}}\xspace}

\def\pythia     {\mbox{\textsc{Pythia}}\xspace}

\def\tell1  {TELL1\xspace}
\def\ukl1   {UKL1\xspace}

\usepackage{cite} 
\usepackage{mciteplus}

\usepackage{longtable} 

\begin{document}

\renewcommand{\thefootnote}{\fnsymbol{footnote}}
\setcounter{footnote}{1}

%
%
\def\Bcand        {\mbox{\B\ candidate}\xspace}
\def\Bcands       {\mbox{\B\ candidates}\xspace}
\def\Bpcand       {\mbox{\Bp\ candidate}\xspace}
\def\Bpcands      {\mbox{\Bp\ candidates}\xspace}
\def\qt           {\mbox{$q_{\mathrm{ T}}$}\xspace}
\def\mhhh        {\ensuremath{m_{h^+h^{\pm}h'^{\mp}}}\xspace}
\def\BsToDphi     {\decay{\Bs}{\Dzb\phi}}
\def\BsToDKstar   {\decay{\Bs}{\Dzb\Kstarzb}}
\def\BdToDKstar   {\decay{\Bd}{\Dzb\Kstarz}}

\def\BtoKpKpPim  {\ensuremath{\Bp\to\Kp\Kp\pim}}  
\def\BtoKpKmPip  {\ensuremath{\Bp\to\Kp\Km\pip}}  
\def\BtoKKPi     {\ensuremath{\Bp\to\Kp\Kpm\pimp}}  

\def\BtoPipPipKm {\ensuremath{\Bp\to\pip\pip\Km}}  
\def\BtoPipPimKp {\ensuremath{\Bp\to\pip\pim\Kp}}  
\def\BtoPiPiK    {\ensuremath{\Bp\to\pip\pipm\Kmp}}  

\def\Etamcsup   {\ensuremath{\eta_{\rm sup}^{MC}}}
\def\Etamcunsup {\ensuremath{\eta_{\rm unsup}^{MC}}}

\def\Etadatasup   {\ensuremath{\eta_{\rm sup}}}
\def\Etadataunsup {\ensuremath{\eta_{\rm unsup}}}

\def\EffMCKKPiOS  {\ensuremath{0.597\pm0.007}}
\def\EffMCKKPiSS  {\ensuremath{0.599\pm0.007}}
\def\EffMCKPiPiOS {\ensuremath{0.977\pm0.007}}
\def\EffMCKPiPiSS {\ensuremath{1.006\pm0.007}}

\def\EffCorKKPiOS  {\ensuremath{46.247\pm0.005}}
\def\EffCorKKPiSS  {\ensuremath{46.100\pm0.005}}
\def\EffCorKPiPiOS {\ensuremath{69.859\pm0.003}}
\def\EffCorKPiPiSS {\ensuremath{70.052\pm0.005}}

\def\EffDataKKPiOS  {\ensuremath{0.276\pm0.003}}
\def\EffDataKKPiSS  {\ensuremath{0.276\pm0.003}}
\def\EffDataKPiPiOS {\ensuremath{0.683\pm0.005}}
\def\EffDataKPiPiSS {\ensuremath{0.705\pm0.005}}

\def\EffRatioKKPi  {\ensuremath{0.9999\pm0.0165}}
\def\EffRatioKPiPi {\ensuremath{0.9685\pm0.0097}}
\def\EffRatioKKPiRnd  {\ensuremath{1.00\pm0.02}}
\def\EffRatioKPiPiRnd {\ensuremath{0.97\pm0.01}}

\def\Ndzbark {\ensuremath{854\pm32}}
\def\Ndzbarpi {\ensuremath{24\, 850\pm170}}

\def\Effdzbark  {\ensuremath{(0.060\pm0.002)\%}}
\def\Effdzbarpi {\ensuremath{(0.097\pm0.003)\%}}

\def\NKPiPiOS {\ensuremath{24\, 044\pm193}}
\def\NKKPiOS  {\ensuremath{955\pm75}}
\def\NKPiPiSS {\ensuremath{2.71\pm9.98}}
\def\NKKPiSS  {\ensuremath{-7.19\pm4.61}}
\def\NKPiPiSSrnd {\ensuremath{2.7\pm10.0}}
\def\NKKPiSSrnd  {\ensuremath{-7.2\pm4.6}}
%

\def\ScaleKKPi  {\ensuremath{5.23\pm0.42}}
\def\ScaleKPiPi {\ensuremath{2.05\pm0.03}}

\def\ULKKPiNinetyZero  {\ensuremath{5.7\times 10^{-8}}}
\def\ULKKPiNinetyFive  {\ensuremath{7.1\times 10^{-8}}}
\def\ULKPiPiNinetyZero {\ensuremath{3.9\times 10^{-8}}}
\def\ULKPiPiNinetyFive {\ensuremath{4.7\times 10^{-8}}}

\def\ULKKPiNinetyZeroFitted  {\ensuremath{1.1\times 10^{-8}}}
\def\ULKKPiNinetyFiveFitted  {\ensuremath{1.8\times 10^{-8}}}
\def\ULKPiPiNinetyZeroFitted {\ensuremath{4.6\times 10^{-8}}}
\def\ULKPiPiNinetyFiveFitted {\ensuremath{5.7\times 10^{-8}}}

\def\KKPiFitBiasValue {\ensuremath{1.05}}
\def\KKPiFitBias      {\ensuremath{0.05}} 
\def\KKPiFitBiasStat  {\ensuremath{0.10}}

\def\KPiPiFitBiasValue {\ensuremath{0.52}}
\def\KPiPiFitBias      {\ensuremath{-0.48}} 
\def\KPiPiFitBiasStat  {\ensuremath{0.15}}

\def\KKPiErrorYinter  {\ensuremath{4.76}}
\def\KPiPiErrorYinter {\ensuremath{9.51}}
\def\KKPiErrorGrad    {\ensuremath{0.09}}
\def\KPiPiErrorGrad   {\ensuremath{0.07}}

\def\KKPiFitBiasYinter  {\ensuremath{0.02}}
\def\KPiPiFitBiasYinter {\ensuremath{-0.56}}
\def\KKPiFitBiasGrad    {\ensuremath{0.03}}
\def\KPiPiFitBiasGrad   {\ensuremath{0.04}}

\def\BFKKPiOS  {\ensuremath{5.0\pm0.7}} 
\def\BFKPiPiOS {\ensuremath{51.0\pm2.9}} 

\def\BFKKPiSS         {\ensuremath{-3.76\pm2.43}} 
\def\BFKPiPiSS        {\ensuremath{5.58\pm20.5}} 
\def\BFKKPiSSrnd         {\ensuremath{-3.8\pm2.4}} 
\def\BFKPiPiSSrnd        {\ensuremath{5.6\pm21.0}} 

\def\BFKKPiRatio         {\ensuremath{-7.5\pm4.9\pm1.0}} 
\def\BFKPiPiRatio        {\ensuremath{1.1\pm4.0\pm0.1}} 

\def\TripleRatio       {\ensuremath{1.00\pm0.08}}
\def\TripleRatioDzbar  {\ensuremath{0.96\pm0.10}}

%
\def\KKPiPidSystPer      {\ensuremath{2.5}} 
\def\KKPiFitBiasSystPer  {\ensuremath{1.4}}
\def\KKPiSystNonBFPer    {\ensuremath{13.0}}
\def\KKPiSystTotalPer    {\ensuremath{19.1}}

\def\KPiPiPidSystPer     {\ensuremath{1.9}}
\def\KPiPiFitBiasSystPer {\ensuremath{10.5}}
\def\KPiPiSystNonBFPer   {\ensuremath{12.1}}
\def\KPiPiSystTotalPer   {\ensuremath{13.3}}

\def\KKPiFitBiasSystEvt  {\ensuremath{0.10}} 
\def\KKPiSystBFEvt       {\ensuremath{1.01}}
\def\KKPiSystNonBFEvt    {\ensuremath{0.94}}
\def\KKPiSystTotalEvt    {\ensuremath{1.38}}
\def\KPiPiFitBiasSystEvt {\ensuremath{0.28}} 
\def\KPiPiSystBFEvt      {\ensuremath{0.15}}
\def\KPiPiSystNonBFEvt   {\ensuremath{0.33}}
\def\KPiPiSystTotalEvt   {\ensuremath{0.36}}

\def\KKPiSystDalitzBF   {\ensuremath{0.15}} 
\def\KKPiSystBFBF       {\ensuremath{0.53}}
\def\KKPiSystBFBFrnd    {\ensuremath{0.5}} 
\def\KKPiPidSystBF      {\ensuremath{0.09}} 
\def\KKPiMCStatsSystBF  {\ensuremath{0.04}}
\def\KKPiFitBiasSystBF  {\ensuremath{0.05}}
\def\KKPiFitModelSystBF {\ensuremath{0.45}} 
\def\KKPiSystNonBFBF    {\ensuremath{0.49}}
\def\KKPiSystNonBFBFrnd {\ensuremath{0.5}} 
\def\KKPiSystTotalBF    {\ensuremath{0.72}}

\def\KPiPiSystDalitzBF   {\ensuremath{0.22}} 
\def\KPiPiSystBFBF       {\ensuremath{0.32}} 
\def\KPiPiSystBFBFrnd    {\ensuremath{0.3}} 
\def\KPiPiPidSystBF      {\ensuremath{0.11}} 
\def\KPiPiMCStatsSystBF  {\ensuremath{0.03}}
\def\KPiPiFitBiasSystBF  {\ensuremath{0.58}} 
\def\KPiPiFitModelSystBF {\ensuremath{0.29}} 
\def\KPiPiSystNonBFBF    {\ensuremath{0.69}}
\def\KPiPiSystNonBFBFrnd {\ensuremath{0.7}} 
\def\KPiPiSystTotalBF    {\ensuremath{0.76}}


\begin{titlepage}
\pagenumbering{roman}

\vspace*{-1.5cm}
\centerline{\large EUROPEAN ORGANIZATION FOR NUCLEAR RESEARCH (CERN)}
\vspace*{1.5cm}
\noindent
\begin{tabular*}{\linewidth}{lc@{\extracolsep{\fill}}r@{\extracolsep{0pt}}}
\ifthenelse{\boolean{pdflatex}}
{\vspace*{-2.7cm}\mbox{\!\!\!\includegraphics[width=.14\textwidth]{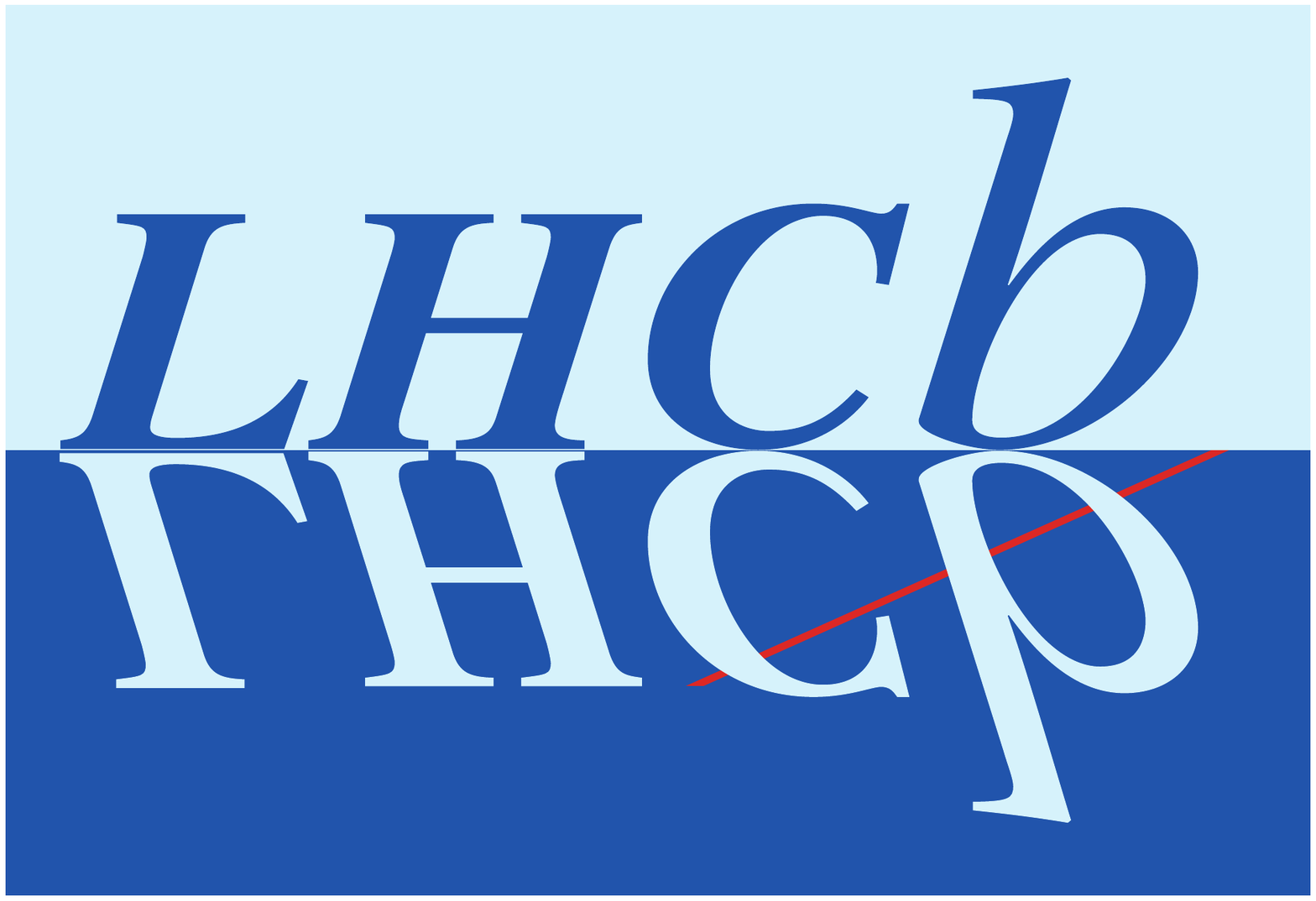}} & &}%
{\vspace*{-1.2cm}\mbox{\!\!\!\includegraphics[width=.12\textwidth]{lhcb-logo.eps}} & &}%
\\
 & & CERN-EP-2016-177 \\  
 & & LHCb-PAPER-2016-023 \\  
 & & 2 December 2016 \\
 & & \\
\end{tabular*}

\vspace*{4.0cm}

{\normalfont\bfseries\boldmath\huge
\begin{center}
 Search for the suppressed decays $\Bp\to\Kp\Kp\pim$ and $\Bp\to\pip\pip\Km$
\end{center}
}

\vspace*{2.0cm}

\begin{center}
The LHCb collaboration\footnote{Authors are listed at the end of this paper.}
\end{center}

\vspace{\fill}

\begin{abstract}
  \noindent
A search is made for the highly-suppressed B meson decays 
$\Bp\to\Kp\Kp\pim$ and $\Bp\to\pip\pip\Km$ using a data sample
corresponding to an integrated luminosity of 3.0\invfb collected by
the \lhcb experiment in proton-proton collisions at centre-of-mass
energies of 7 and 8\tev. No evidence is found for the decays, and upper
limits at 90\% confidence level are determined to be
$\BF(\Bp\to\Kp\Kp\pim) < \ULKKPiNinetyZeroFitted$ and
$\BF(\Bp\to\pip\pip\Km) < \ULKPiPiNinetyZeroFitted$.
\end{abstract}

\vspace*{2.0cm}

\begin{center}
  Published in Phys. Lett. B 765 (2017) 307
\end{center}

\vspace{\fill}

{\footnotesize 
\centerline{\copyright~CERN on behalf of the \lhcb collaboration, licence \href{http://creativecommons.org/licenses/by/4.0/}{CC-BY-4.0}.}}
\vspace*{2mm}

\end{titlepage}


\newpage
\setcounter{page}{2}
\mbox{~}

\cleardoublepage


\renewcommand{\thefootnote}{\arabic{footnote}}
\setcounter{footnote}{0}



\pagestyle{plain} 
\setcounter{page}{1}
\pagenumbering{arabic}


%

\section{Introduction}
\label{sec:Introduction}

\begin{figure}[ht]
\begin{center}
\hspace*{-3.5mm}
\begin{tabular}{cc}
  \includegraphics[width=0.49\columnwidth]{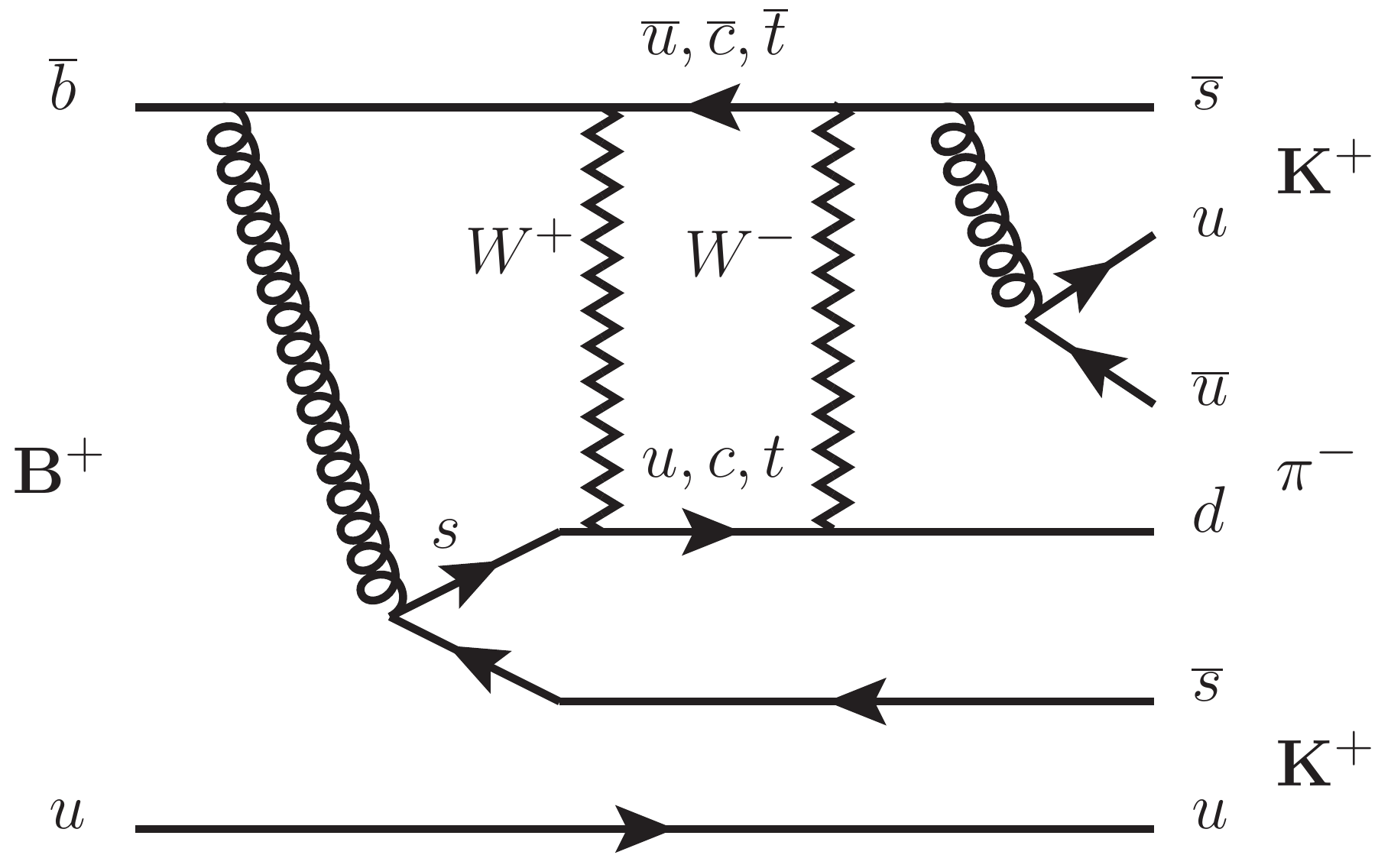} &
  \includegraphics[width=0.49\columnwidth]{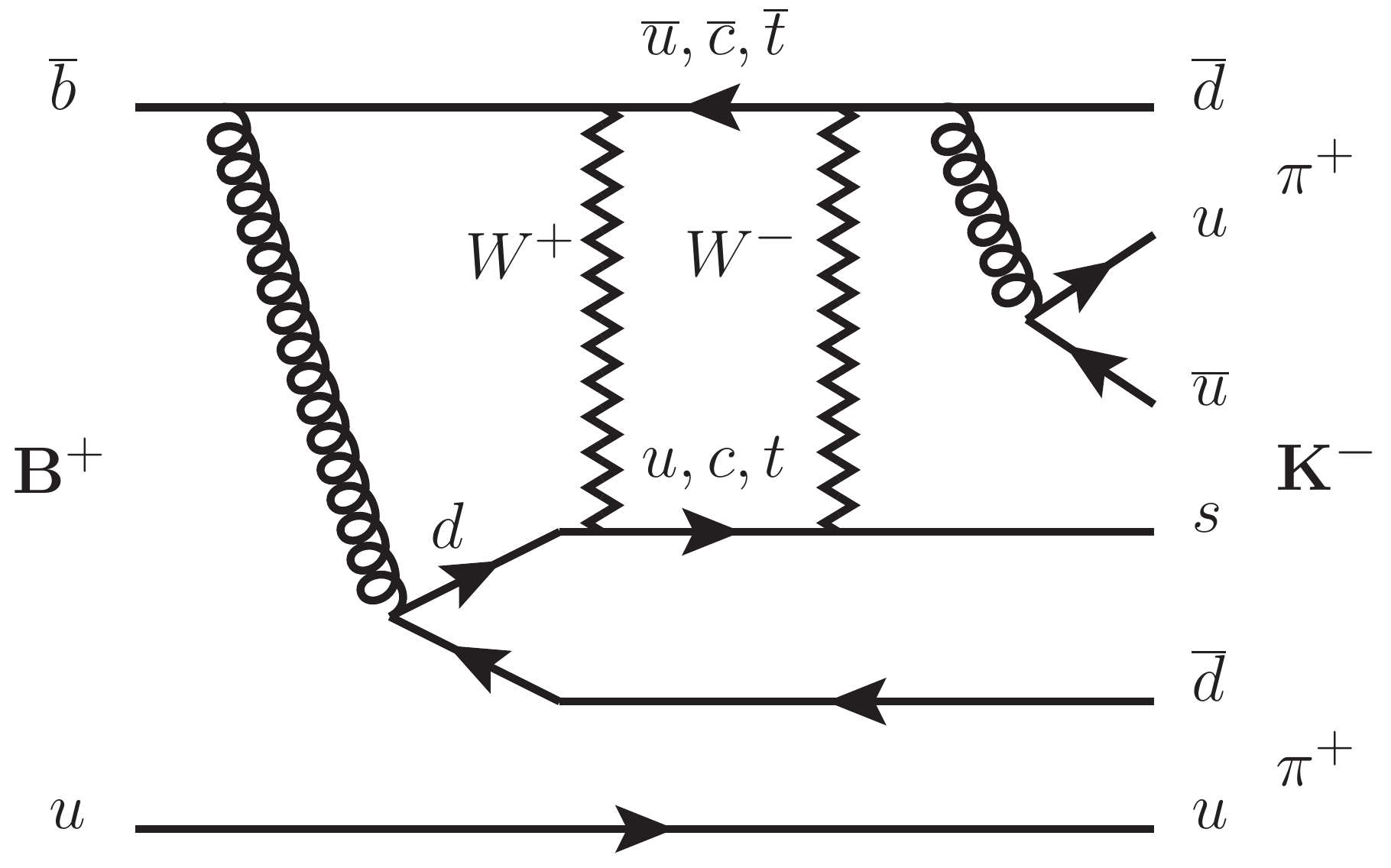} \\
\end{tabular}
\end{center}
\caption{Main SM 
  diagrams for the suppressed decays (left) \BtoKpKpPim\  and
  (right) \BtoPipPipKm.}
\label{fig:box}
\end{figure}

Transitions of the type $\bquark\to\squark\squark\dquarkbar$ and
$\bquark\to\dquark\dquark\squarkbar$ are rare in the Standard Model
(SM)~\cite{Huitu:1998vn,Pirjol:2009vz}. The calculation of the
$\bquark\to\squark\squark\dquarkbar$ amplitude results in branching
fractions of at most $\mathcal{O}(10^{-11})$, the exact value
depending on the unknown relative phase between \tquark\ and
\cquark\ quark contributions in the \Wpm-exchange
box~\cite{Fajfer:2000ny}, as shown in Fig.~\ref{fig:box}. The magnitude
of the $\bquark\to\dquark\dquark\squarkbar$ amplitude is expected to
be even smaller due to the relative $|\Vtd/\Vts|$ factor, leading to
predicted branching fractions of
$\mathcal{O}(10^{-14})$~\cite{Fajfer:2006av}.

Physics beyond the Standard Model (BSM) could result in enhanced
branching fractions that can be detected at current
experiments. Theoretical models that have been investigated include
the Minimal Supersymmetric Standard Model with and without
R-Parity Violation, variations of the two Higgs doublet model,
and extensions of the SM where an additional flavour changing
$\Z^{\prime}$ neutral boson is
present~\cite{Fajfer:2000ny,Fajfer:2006av}. In these SM extensions,
for certain plausible values of the model parameters,
$\bquark\to\squark\squark\dquarkbar$ and
$\bquark\to\dquark\dquark\squarkbar$ transitions may lead to branching
fractions of up to $10^{-8}$ and $10^{-7}$, respectively. It has been
suggested by these theoretical studies that the most suitable
three-body decay modes to see the effects of BSM physics in such
transitions are the \BtoKpKpPim\ and \BtoPipPipKm\ decays,
where two particles in the final state have the same
flavour and charge.\footnote{The inclusion of charge-conjugate decays
  is implied throughout.}

An upper limit of \ensuremath{1.29\times 10^{-4}} at 90\% confidence level
on the branching fraction for the \BtoPipPipKm\ decay
was first determined by OPAL~\cite{Abbiendi:1999st}. Improvements in
sensitivity were obtained by the \epem\ $B$-factories, and currently the 90\%
confidence level upper limits for \BtoKpKpPim\ and
\BtoPipPipKm\ decays are $1.6\times 10^{-7}$ and $9.5\times 10^{-7}$,
respectively~\cite{Abe:2002av,Aubert:2003xz,Aubert:2008rr}.

This paper reports on the search for the suppressed decays
\BtoKpKpPim\ and \BtoPipPipKm\ using data samples corresponding to
1.0 and 2.0\invfb collected by LHCb at \sqs = 7 and 8\tev,
respectively. The corresponding unsuppressed decays \BtoKpKmPip\ and
\BtoPipPimKp\ are used for normalisation.

\section{Detector and simulation}
\label{sec:detector}

The \lhcb detector~\cite{Alves:2008zz,LHCb-DP-2014-002} is a
single-arm forward spectrometer covering the \mbox{pseudorapidity}
range $2<\eta<5$, designed for the study of particles containing
\bquark\ or \cquark quarks. The detector includes a high-precision
tracking system consisting of a silicon-strip vertex detector
surrounding the $pp$ interaction region, a large-area silicon-strip
detector located upstream of a dipole magnet with a bending power of
about $4{\mathrm{\,Tm}}$, and three stations of silicon-strip
detectors and straw drift tubes placed downstream of the magnet.  The
polarity of the dipole magnet is reversed periodically throughout
data-taking. The configuration with the magnetic field vertically
upwards (downwards) bends positively (negatively) charged particles in
the horizontal plane towards the centre of the LHC. The tracking
system provides a measurement of momentum, \ptot, of charged particles
with a relative uncertainty that varies from 0.5\% at low momentum to
1.0\% at 200\gevc. The \mbox{impact parameter (IP)} is the minimum
distance of a track to a primary vertex (PV) and is measured with a
resolution of $(15+29/\pt)\mum$, where \pt is the component of the
momentum transverse to the beam, in\,\gevc.  Different types of
charged hadrons are distinguished using information from two
ring-imaging Cherenkov (\rich) detectors~\cite{LHCb-DP-2012-003}.
Photons, electrons and hadrons are identified by a calorimeter system
consisting of scintillating-pad and preshower detectors, an
electromagnetic calorimeter and a hadronic calorimeter. Muons are
identified by a system composed of alternating layers of iron and
multiwire proportional chambers~\cite{LHCb-DP-2012-002}.


The online event selection is performed by a
trigger~\cite{LHCb-DP-2012-004}, which consists of a hardware stage,
based on information from the calorimeter and muon systems, followed
by a software stage, which applies a full event reconstruction. At the
hardware stage, the candidates are triggered either one
of the particles from the \bquark\ candidate decay depositing a
transverse energy of at least $3500$\mev\ in the calorimeter, or by
other activity in the event, mainly associated with the decay products
of the other \bquark\ hadron produced in the $pp$ primary
interaction. 

Simulated \BtoKKPi\ and \BtoPiPiK\ decays, generated uniformly in
phase space, are used to optimize the suppressed signal selections
and to evaluate the ratios of the efficiencies for each suppressed decay mode relative
to their corresponding unsuppressed decay modes.
In the simulation, \proton\proton collisions are generated using \pythia
8~\cite{Sjostrand:2006za,Sjostrand:2007gs} with a specific
\lhcb\ configuration~\cite{LHCb-PROC-2010-056}. Decays of hadronic
particles are described by \evtgen~\cite{Lange:2001uf} in which final
state radiation is generated by \photos~\cite{Golonka:2005pn}. The
interaction of the generated particles with the detector and its
response are implemented using the \geant
toolkit~\cite{Allison:2006ve} as described in
Ref.~\cite{LHCb-PROC-2011-006}.

\section{Event selection}
\label{sec:selection}


The candidate \BtoKKPi\ and \BtoPiPiK\ decays are reconstructed using
three charged tracks with mass hypotheses and total charge consistent
with the decay.  The final state particles are required to have a good
track fit with a reduced chi-square $\chisqndf < 3$. All three tracks
must have momentum $\ptot > 1500\mevc$, transverse momentum $\pt >
100\mevc$, and $\chisqip > 1$ with respect to all PVs in the
event. The quantity \chisqip is defined as the difference between the
vertex-fit \chisq\ of the PV reconstructed with and without the
considered track. Combinatorial backgrounds are suppressed by
requiring that the scalar sum of the \pt\ of the tracks is greater
than $4500\mevc$ and the sum of the tracks' $\chisqip >200$.  The
track with the highest \pt\ must have IP $>$ 0.05\mm. The second
highest track \pt\ is required to be greater than $900$\mevc. The
maximum distance of closest approach between tracks has to be less
than 0.2\mm.


The information from the \rich, calorimeter and muon systems is used
for particle identification (PID). Muons are rejected by a veto
applied to each track. Loose kaon and pion PID is required for the
remaining charged tracks to reduce the number of wrong combinations
before forming a \Bpcand.

The reconstructed \Bpcands\ are required to have an invariant mass in
the range $5080 - 5480\mevcc$, $\pt>1700\mevc$, $\chisqip<10$, vertex
fit $\chisqndf < 12$, distance between the PV and the decay point (or
secondary vertex, SV) greater than 3\mm, and a significant
displacement between primary and secondary vertex, with the three
dimensional \chisq-distance between the two greater than 700.  When
more than one PV is reconstructed, the one with the minimum
\chisqip\ for the \Bpcand\ is chosen. The cosine of the angle
$\theta_f$ between the reconstructed momentum of the \Bpcand\ and the
\Bpcand\ flight direction is required to be $\cos\theta_f >
0.99998$. The pointing variable $\theta_p \equiv
P_B\sin\theta_f/(P_B\sin\theta_f+\sum_i^3\qt_i)$ is required to be
less than $0.05$, where $P_B$ is the total momentum of the
three-particle final state and $\sum_i^3\qt_i$ is the sum of the
transverse momenta of the three tracks with respect to the momentum
direction of the \Bpcand. These requirements remove additional
combinatorial background and partially reconstructed \bquark\ hadron
decays.

Depending on its charge and the polarity of the dipole magnet, a
charged particle traversing the magnetic field can be bent
horizontally into or out of the detector acceptance. To minimise
charge-dependent differences in the reconstruction efficiencies for
the signal or normalisation channels caused by the magnetic field, an
additional criterion is placed on the $x$ and $z$
components\footnote{In the LHCb right handed coordinate system, the
  $z$-axis points from the interaction point into the experiment and
  the $y$-axis is vertical, pointing upwards.} of the
\Bpcand\ momentum such that $|p_x| \leq 0.317 \times (p_z -
2400\mevc)$~\cite{HamishGordon:2013}, which ensures that tracks of
both charges are well contained in the detector acceptance.


A \Bpcand\ is rejected if the invariant mass formed from two of the
charged tracks with opposite charge is within 25\mevcc of the
\Dz\ mass. The \Dz veto suppresses possible background from
         {\ensuremath{\Bp\to\Dz h^{+}}} decays. Backgrounds from
         {\ensuremath{\Bz\to\Dm h^{+}}} decays are found to be
         negligible.  For \BtoPipPimKp\ decays only, an additional
         invariant mass criterion of $|3104\mevcc -
         m_{\pip\pim}|>20$\mevcc is required to eliminate
         contamination $\jpsi\to\mumu$, where the muons are
         misidentified as pions, and from $\jpsi\to\pip\pim$.


The reconstructed candidates that meet the above criteria are filtered
using a boosted decision tree (BDT) algorithm~\cite{Breiman,AdaBoost}. The BDT
is trained with a sample of simulated signal candidates and a
background sample of data candidates taken from the \Bpcand\ invariant
mass sideband above $5400$\mevcc, which is dominated by combinatorial
background. The training is performed separately
for events that have been triggered by information from the signal
decay only and for events that have been triggered by other
particles.  This is required as the two trigger scenarios have different
sensitivities to the signal and background. The input variables are
chosen to produce the best discrimination and to minimise the
dependence of the BDT on the mass of the \Bpcand, the PID variables
and on the kinematic configuration of the \Bpcand parametrised in the
Dalitz plane.
The twelve variables used by the BDT
are: the \Bpcand \pt; the flight distance between the PV and
\Bpcand\ SV; the angle $\theta_f$; the
\Bpcand\ pointing angle $\theta_p$; the \pt\ of the track
with the lowest \pt; the sum of the tracks' \pt; the sum of the 
\chisqip\ of the three tracks; the IP of
the track with the highest \pt, with respect to the PV; the
momentum \ptot\ of each of the three tracks; and the \chisqip of
the track reconstructed with the \pion\ hypothesis for \BtoKKPi\ decays
or the \kaon\ hypothesis for \BtoPiPiK\ decays. The same set of
variables is found to result in a robust multivariate classifier for the
four decay channels under consideration.


A figure of merit ${\rm FOM} \equiv \epsilon/(0.5\times N_{\sigma} +
\sqrt{N_B})$, is used to identify the optimal BDT selection
criteria. Here $\epsilon$ is the simulated signal selection
efficiency, $N_B$ is the number of background events that pass the
selection and have a mass in a $\pm 50\mevcc$ window around the \Bpm
mass~\cite{PDG2014}, and $N_{\sigma}$ is the required significance,
expressed in terms of standard deviations from the hypothesis of a
null signal~\cite{Punzi:2003bu}. The quantity FOM is optimized with
$N_{\sigma}=5$, but the final result is robust against changes of one
or two units in $N_{\sigma}$. For events that pass the selection
criteria described above, the optimal BDT selection criterion for
\BtoKpKpPim\ results in 85\% of the simulated signal events being
accepted and 71\% of the background events rejected. For \BtoPipPipKm,
67\% of the simulated signal events are accepted and 91\% of the
background events are rejected.

After the BDT selection criterion has been applied, each track is
required to pass PID criteria. Each track has a probability to be a
kaon and a probability to be a pion, leading in total to six possible
PID assignments for a \Bpcand. The same FOM optimisation described
above is performed for each of the six cases in turn, starting with
the track with the highest \pt. After the application of these
criteria, less than 2--4\% have more than one candidate, depending on
the decay mode. For these multiple candidate events, one candidate is
chosen at random and the others discarded.

The efficiencies of all the selection requirements are calculated with
simulated events. The PID efficiency for hadrons is determined from
data using large calibration samples of
$\Dstarp\to\Dz(\to\Km\pip)\pip$ decays. The PID efficiencies in the
simulated sample are corrected by reweighting the calibration sample
to match the momentum and the pseudorapidity distributions of the
final state particles in the signal decay and the multiplicity of
tracks in the event. The effective kaon and pion PID
  efficiencies in this analysis are approximately $90\%$ and $80\%$,
  respectively. The rates for misidentifying a kaon as a pion or a
  pion as a kaon are less than 0.1\%.

After all selection criteria have been applied, the ratio of the
\BtoKpKmPip\ to \BtoKpKpPim\ selection efficiencies, weighted by the
integrated luminosity taken with different magnet polarities and beam
energy, is \EffRatioKKPiRnd, while the ratio of the \BtoPipPimKp\ to
\BtoPipPipKm\ selection efficiencies is \EffRatioKPiPiRnd. The quoted
uncertainties are statistical only and are due to the limited
simulation sample size. For each of the four modes, the
  selection efficiency across the Dalitz plane is constant with a
  relative variation of less than 9\%.

\section{Determination of the signal yields}
\label{sec:yield}


Signal yields are determined from simultaneous, unbinned and extended
maximum likelihood fits to the invariant mass distributions \mhhh\ of
the suppressed and unsuppressed decays, where $h$ and $h'$ denote \pion\ or
\kaon. Separate fits are made for \BtoKpKpPim\ and
\BtoKpKmPip\ decays, and for \BtoPipPipKm\ and \BtoPipPimKp\ decays.

The signal component for both the suppressed and unsuppressed decays is
parameterised as the sum of a Gaussian function and a Crystal Ball
function~\cite{Skwarnicki:1986xj}. The values of the signal function
parameters are constrained to be the same for both the suppressed and
unsuppressed signal components. 

For all decay modes, the combinatorial background is
parameterised with an exponential function. Partially
reconstructed backgrounds, largely due to \B\ decays with four
or more particles in the final state, where one or more particles are
not reconstructed, appear predominantly at \mhhh\ masses below
5150\mevcc\ and are modelled with an \argus\
function~\cite{Albrecht:1990cs} convolved with a Gaussian resolution
function. In the case of the suppressed \BtoPipPipKm\ mode, an
additional component, modelled in the same way, is used to account for \Bs\ four-body
decays such as $\Bs\to\Dsm\pip$, with $\Dsm\to\Km\pip\pim$, where one of the
decay particles is not reconstructed. These appear at \mhhh\ masses below
5250\mevcc. The slope parameter of this \argus\ function is fixed from a
fit to simulated decays. The signal yields, background yields, and all
signal and background parameters (apart from the fixed slope parameter of the
$\Bs\to\Dsm\pip$ component) are allowed to float in the fit.

To investigate the presence of any peaking background in the signal
mass region, a total of 350 million events from 200 \B, \Bs,
\Lb, and \Xib\ decays are simulated and reconstructed using the same
selection criteria used for the signal decay modes. In addition, the
data are divided into two samples, above and below the signal mass
region $5230 < \mhhh < 5320\mevcc$. The Dalitz plot distributions in
each sample are studied to identify any resonances that might be
present in the signal region. There is no evidence for any peaking
background in the signal mass region.

The performance of the fit procedure is tested by creating ensembles
of simulated datasets generated from the functions fitted to the
data. A large number of datasets is generated and fits are performed
with the number of signal and background events left free to fluctuate
according to a Poisson distribution. The fit biases on the signal
yields extracted from the pseudoexperiments are
$\KKPiFitBias\pm\KKPiFitBiasStat$ and
$\KPiPiFitBias\pm\KPiPiFitBiasStat$ events for the \BtoKpKpPim\ and
\BtoPipPipKm\ decays, respectively, where the uncertainty is
statistical only.

\begin{figure}[ht!]
\begin{center}
\begin{tabular}{c}
  \includegraphics[width=0.78\columnwidth]{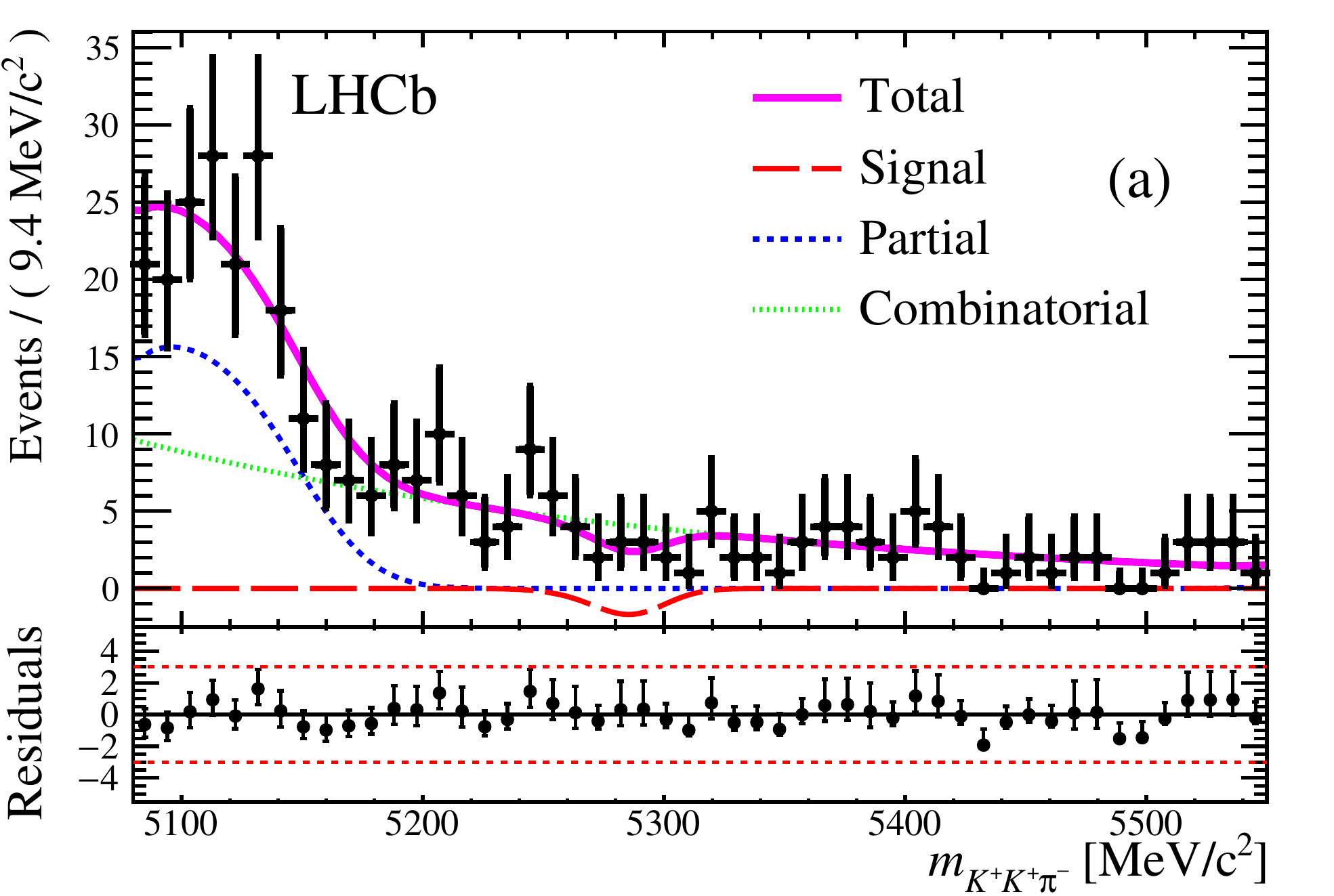} \\
  \includegraphics[width=0.78\columnwidth]{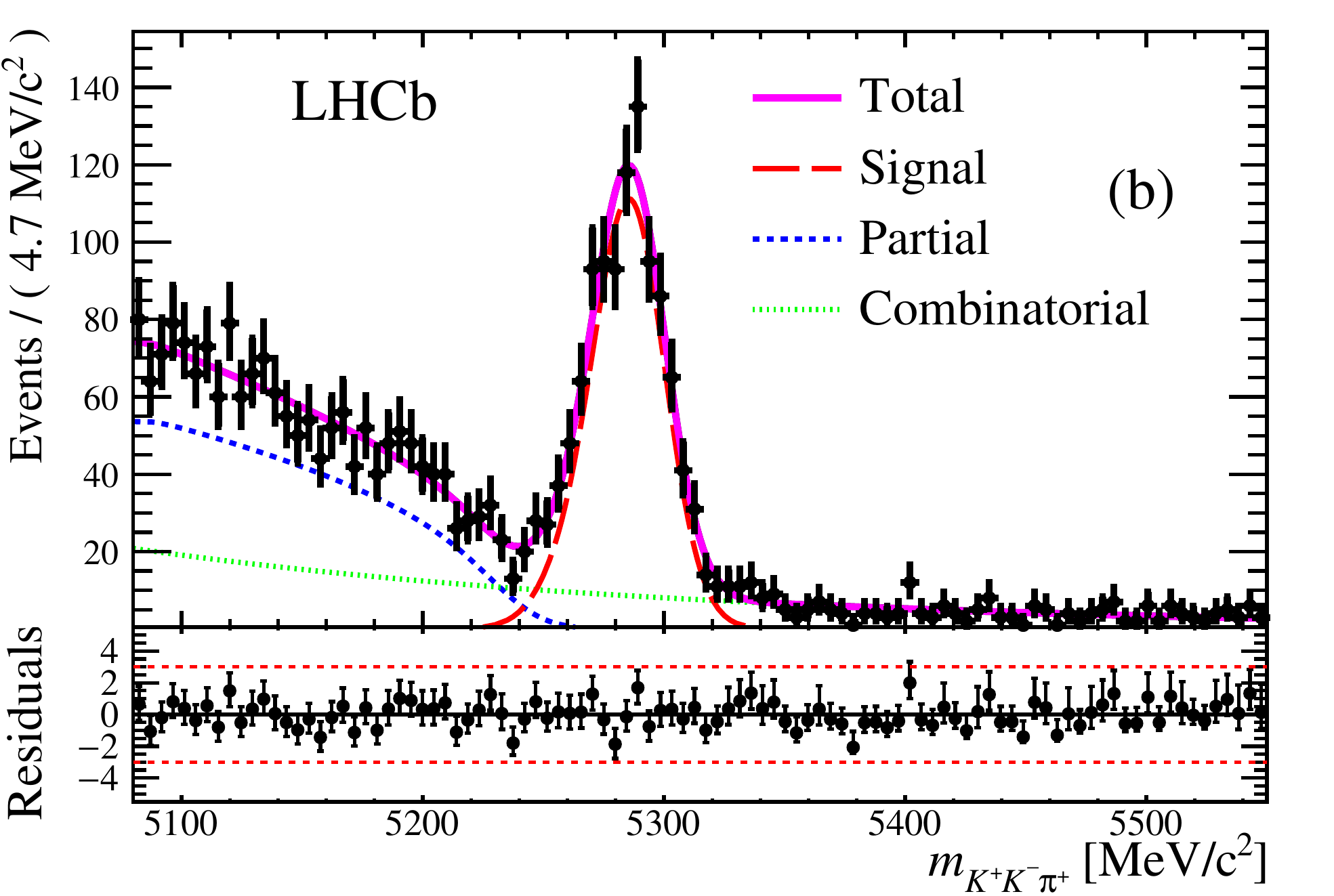}
\end{tabular}
\caption{Invariant mass spectra of (a) \BtoKpKpPim\ and (b)
  \BtoKpKmPip\ candidates, with the results of the unbinned extended maximum
  likelihood fits overlaid. The dashed (blue) line represents the
  partially reconstructed background, the dotted (green) line the
  combinatorial background, the long dashed (red) line is the signal, and the solid
  (magenta) line the total. Residual differences between data and the
  fits are shown  below the mass plots in units of standard deviation.}
\label{fig:KKPi_result}
\end{center}
\end{figure}
\begin{figure}[ht!]
\begin{center}
\begin{tabular}{c}
  \includegraphics[width=0.78\columnwidth]{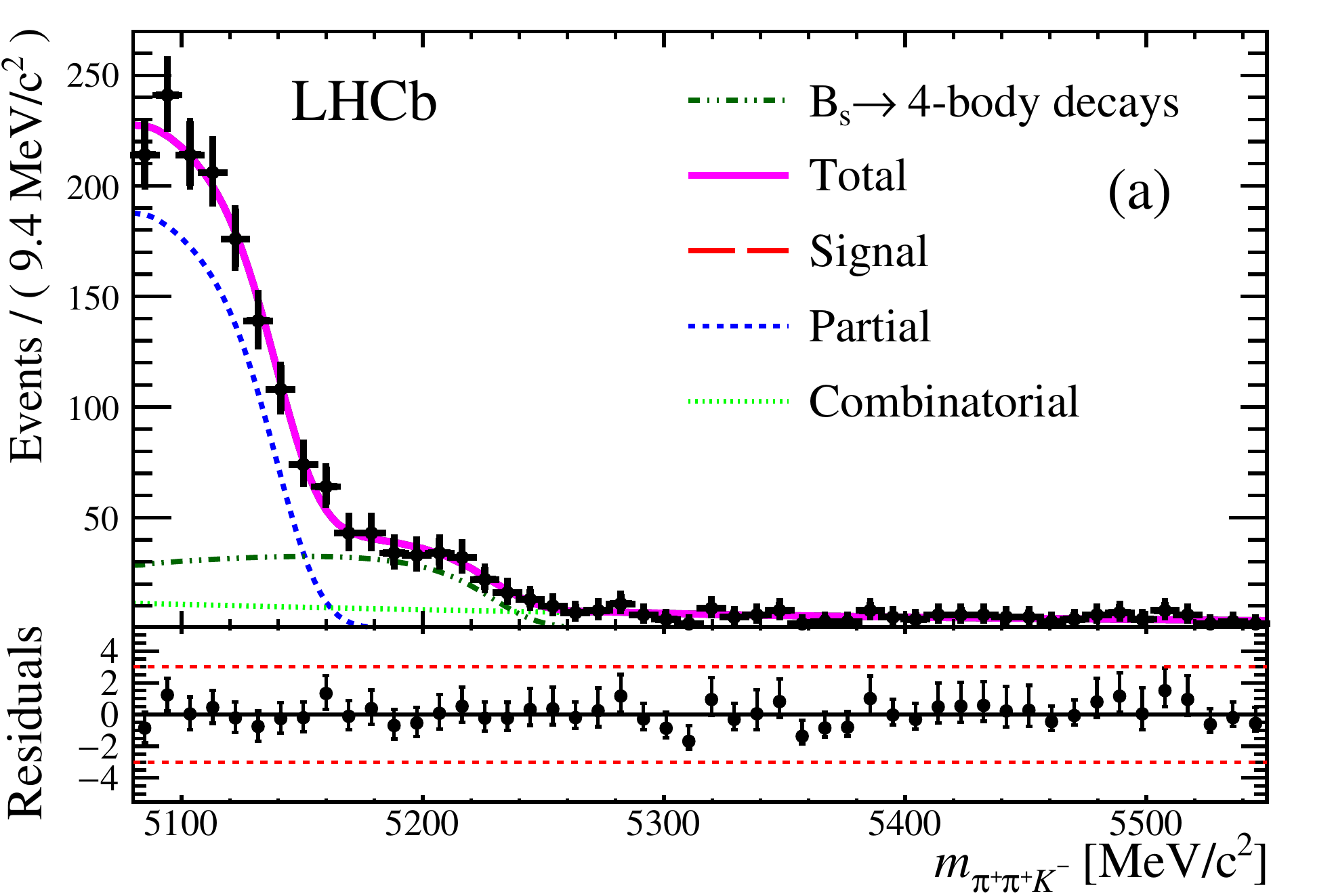} \\
  \includegraphics[width=0.78\columnwidth]{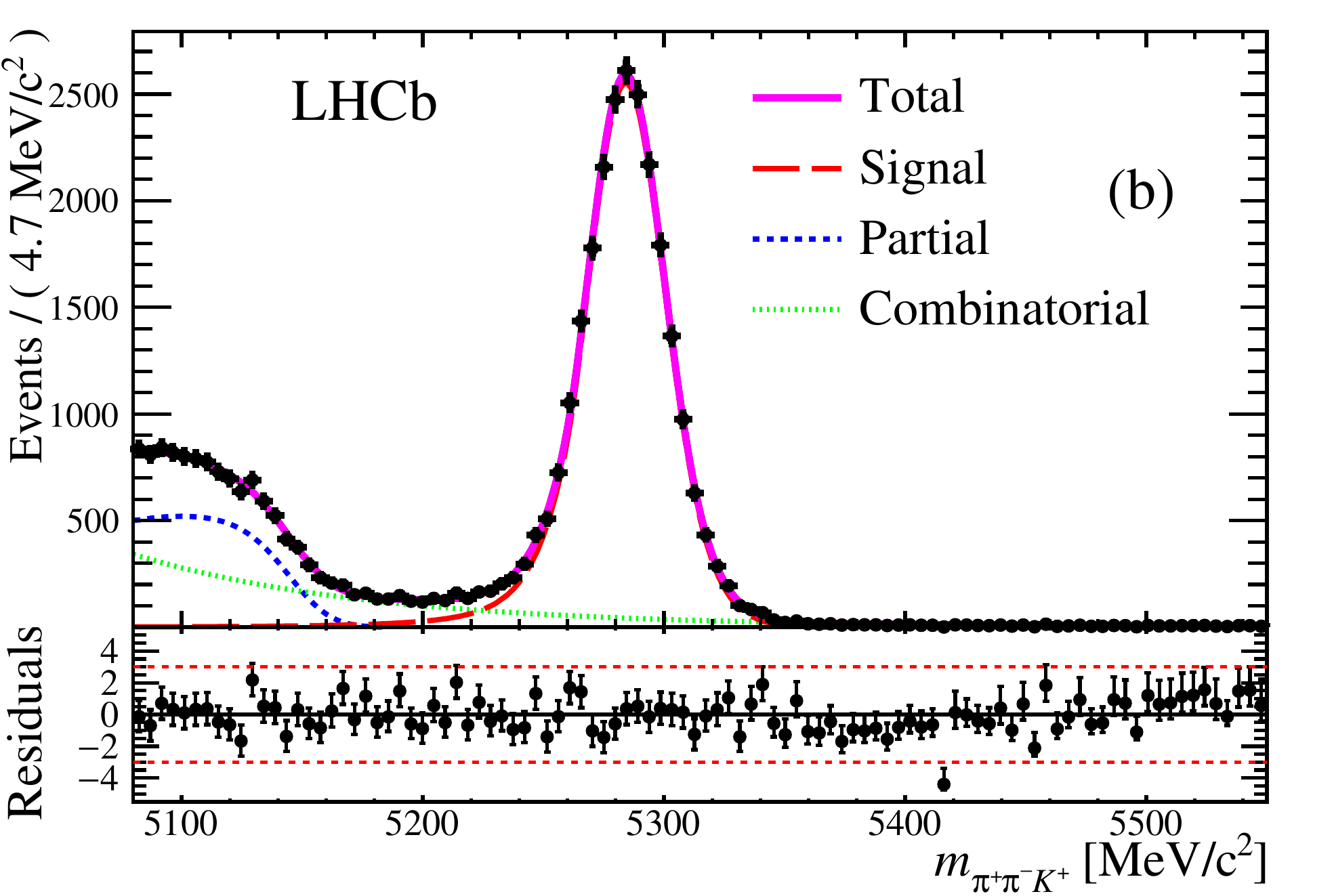}
\end{tabular}
\caption{Invariant mass spectra of (a) \BtoPipPipKm\ and (b)
  \BtoPipPimKp\ candidates, with the results of the unbinned extended maximum
  likelihood fits overlaid. The dashed (blue) line represents the
  partially reconstructed background, the dotted (green) line the
  combinatorial background, the long dashed (red) line is the signal, the
  dot-dash (dark green) line the \Bs\ four-body backgrounds, and the solid
  (magenta) line the total. Residual differences between data and the
  fits are shown below the mass plots in units of standard deviation.}
\label{fig:KPiPi_result}
\end{center}
\end{figure}

Figure~\ref{fig:KKPi_result} shows the fit to the \BtoKpKpPim\ and
\BtoKpKmPip\ decay candidates. The signal yields are \NKKPiSSrnd\ and
\NKKPiOS\ for \BtoKpKpPim\ and \BtoKpKmPip\, respectively.
Figure~\ref{fig:KPiPi_result} shows the fit to the \BtoPipPipKm\ and
\BtoPipPimKp\ decay candidates.  The signal yields are
\NKPiPiSSrnd\ and \NKPiPiOS\ for \BtoPipPipKm\ and \BtoPipPimKp,
respectively.  The \BtoKpKpPim\ and \BtoPipPipKm\ signal yields have
been corrected for the fit bias introduced by the fitting procedure.

The branching fractions of the suppressed decays are calculated using
\begin{equation}
\label{eq:bf}
  {\cal B}_{\rm sup} = \frac{N^{\rm sup}_{\rm sig}}{N^{\rm unsup}_{\rm sig}}
  \times \frac{\epsilon^{\rm unsup}}{\epsilon^{\rm sup}} \times {\cal B}_{\rm unsup},
\end{equation}
\noindent where $N^{\rm unsup}_{\rm sig}$ and $N^{\rm sup}_{\rm sig}$ are the numbers
of fitted signal events for the unsuppressed and suppressed modes
(corrected for fit bias), while $\epsilon^{\rm unsup}$ and
$\epsilon^{\rm sup}$ are the selection efficiencies calculated from simulated
events and corrected for differences in selection efficiency between
simulation and data~\cite{Anderlini:2202412}. Finally, ${\cal B}_{\rm unsup}$ is the known
branching fraction for the unsuppressed reference mode~\cite{PDG2014}.

%
%
\section{Systematic uncertainties}
\label{sec:systematics}


The measurements of the branching fractions of the suppressed modes
depend on the ratios of suppressed to unsuppressed signal yields and
selection efficiencies. Since the final state is the same, apart from
the charge assignment, the ratio of unsuppressed and suppressed
selection efficiencies is close to unity, and most of the systematic
uncertainties cancel.

The main systematic uncertainties on the branching fractions are due
to the unsuppressed branching fraction uncertainties,
PID charge dependence, discrepancies in PID between data and simulated
events, fit biases, alternative mass fit models, simulated event
statistics, and assumptions concerning the Dalitz plot distributions.

The uncertainties on the known \BtoKpKmPip\ and
\BtoPipPimKp\ branching fractions result in systematic uncertainties
of $\KKPiSystBFBF\times 10^{-8}$ and $\KPiPiSystBFBF\times 10^{-9}$,
respectively~\cite{Garmash:2005rv,Aubert:2008bj,Aubert:2007xb}.
Systematic uncertainties of $0.04\times 10^{-8}$ and $0.03\times
10^{-9}$ are assigned to the \BtoKpKmPip\ and \BtoPipPimKp\ decay
modes, respectively, to account for the effect of limited simulated
events available to determine the reconstruction efficiencies.

Studies have shown a small PID dependence on the track charge with
differences in efficiency below $0.2\%$ for $\pipm$ and below $0.4\%$ for
\Kpm. Systematic uncertainties of $0.04\times 10^{-8}$ are assigned to
both the \BtoKpKmPip\ and the \BtoPipPimKp\ branching fractions,
derived from a $0.2\%$ systematic uncertainty per pion and $0.4\%$ per
kaon, added linearly. The PID efficiency is extracted from data and
systematic uncertainties of $\KKPiPidSystBF\times 10^{-8}$ and
$\KPiPiPidSystBF\times 10^{-9}$ are applied to \BtoKpKpPim\ and
\BtoPipPipKm, respectively, to account for different data-taking conditions.

The process used to determine the selection criteria for the BDT output
and the six PID probabilities is validated by changing the order in
which the PID criteria are optimized, adjusting the FOM for 
the predicted suppressed signal yield (using published branching
fraction upper limits~\cite{Abe:2002av,Aubert:2003xz,Aubert:2008rr})
rather than the signal reconstruction efficiency $\epsilon$, and
reweighting the simulation samples to match PID distributions in
data~\cite{Anderlini:2202412}. The process is shown to be robust and
no systematic uncertainty is applied.

To allow for possible differences in reconstruction efficiency as a
function of position in the Dalitz plane due to resonances and
interference, simulated events are generated with a distribution of
resonances taken from previous published results, which are however
only available for the
\BtoPipPimKp\ decay~\cite{Garmash:2005rv,Aubert:2008bj}. The average
values of the reconstruction efficiencies for the phase-space and
resonance-allowed Dalitz plots differ by $(8\pm2)\%$, which is taken
as the difference between a phase-space and a resonance-allowed
distribution for all modes. The resulting uncertainties are
$\KKPiSystDalitzBF\times 10^{-8}$ and $\KPiPiSystDalitzBF\times
10^{-9}$ for \BtoKpKpPim\ and \BtoPipPipKm, respectively.

The fit bias from the ensemble of simulated pseudoexperiments
generated from the fit to data is used to correct the signal yield
observed in data.  The systematic uncertainty on this correction is
defined as half the fit bias added in quadrature with the fit bias
statistical uncertainty. The systematic uncertainties introduced by
this procedure are $\KKPiFitBiasSystBF\times 10^{-8}$ and
$\KPiPiFitBiasSystBF\times 10^{-9}$ for the \BtoKpKpPim\ and
\BtoPipPipKm\ branching fractions, respectively.

To understand the impact of the fit model on the results, the
components of the default models are changed independently. The signal
component is replaced with a Crystal Ball function. The slope
parameter of the $\Bs\to\Dsm\pip$ \argus\ function is allowed to float
and the $\Bs\to\Dsm\pip$ component is replaced with a bifurcated
Gaussian function. The combinatorial background distribution is
modelled with a second order polynomial instead of an
exponential. 
Studies of cross-feed from simulated unsuppressed $\Bp\to
h^+h^-h^+$ decays, where the flavour of one or more particles is
misidentified, indicate $5.7\pm2.7$ events in the \BtoKpKpPim\ decay
mode and $3.1\pm1.8$ events in \BtoPipPipKm\ decay mode. The
uncertainty is dominated by limited simulation sample size. Cross-feed
events are shifted by a minimum of $\sim40\mevcc$ above and below the
\B\ mass and the majority of the events do not appear in the signal
region. To confirm this, the fit is repeated with two additional
Gaussian functions centred around 5240\mevcc and 5320\mevcc,
respectively. As a cross-check, an additional fit is performed with the means and
widths of the Gaussian component allowed to vary. The fitted yields
are compatible with zero. Systematic uncertainties of
$\KKPiFitModelSystBF\times 10^{-8}$ and $\KPiPiFitModelSystBF\times
10^{-9}$ are assigned for the \BtoKpKpPim\ and \BtoPipPipKm\ decays,
respectively.

The total systematic uncertainty is determined by adding the
individual contributions in quadrature. A summary of the systematic
uncertainties is given in Table~\ref{tab:systematics}. The total
systematic uncertainties are $\KKPiSystTotalBF\times 10^{-8}$ and
$\KPiPiSystTotalBF\times 10^{-9}$ for \BtoKpKpPim\ and
\BtoPipPipKm, respectively.

\begin{table}[htbp!]
\caption[Systematic uncertainties] {Systematic uncertainties on the
 \BtoKpKpPim\ and \BtoPipPipKm\ branching fractions in
 units of $10^{-8}$ and $10^{-9}$, respectively.}
\begin{center}
\begin{tabular}{lcc}
Criteria & \BtoKpKpPim & \BtoPipPipKm \\
\hline 
Simulation statistics    & \KKPiMCStatsSystBF & \KPiPiMCStatsSystBF \\
PID charge dependence    & 0.04 & 0.04 \\
PID discrepancies        & \KKPiPidSystBF & \KPiPiPidSystBF \\
Dalitz plot efficiencies & \KKPiSystDalitzBF & \KPiPiSystDalitzBF \\
Fit bias                 & \KKPiFitBiasSystBF  & \KPiPiFitBiasSystBF \\
Fit model                & \KKPiFitModelSystBF & \KPiPiFitModelSystBF\\
\hline
Subtotal  & \KKPiSystNonBFBF & \KPiPiSystNonBFBF \\
PDG ${\cal B}_{\rm unsup}$ uncertainty~\cite{PDG2014} & \KKPiSystBFBF & \KPiPiSystBFBF \\
Total                 & \KKPiSystTotalBF & \KPiPiSystTotalBF \\
\end{tabular}
\end{center}
\label{tab:systematics}
\end{table}

\section{Results and conclusions}
\label{sec:results}

Including all statistical and systematic uncertainties, the
ratios of branching fractions are calculated to be
\begin{eqnarray}
  \frac{\BF(\BtoKpKpPim)}{\BF(\BtoKpKmPip)}  & = & (\BFKKPiRatio)\times 10^{-3},  \nonumber \\
  \frac{\BF(\BtoPipPipKm)}{\BF(\BtoPipPimKp)} & = & (\BFKPiPiRatio)\times 10^{-4}, \nonumber
\end{eqnarray}

\noindent where the first uncertainties are statistical and the second are
systematic. Using the above and the world average of the unsuppressed
branching fractions~\cite{PDG2014} and using Eq.~\ref{eq:bf}, we calculate the branching fractions

\begin{eqnarray}
  \BF(\BtoKpKpPim)  & = &
  (\BFKKPiSSrnd\pm\KKPiSystNonBFBFrnd\pm\KKPiSystBFBFrnd)\times 10^{-8}, \nonumber \\
  \BF(\BtoPipPipKm) & = &
  (\BFKPiPiSSrnd\pm\KPiPiSystNonBFBFrnd\pm\KPiPiSystBFBFrnd)\times 10^{-9}, \nonumber \\
\end{eqnarray}

\noindent where the first uncertainties are statistical, the second are
systematic without the unsuppressed decay branching fraction
uncertainties, and the third are associated with the present knowledge
of the \BtoKpKmPip\ and \BtoPipPimKp\ branching fractions.

\begin{figure}[ht!]
\begin{center}
\begin{tabular}{c}
  \includegraphics[width=0.78\columnwidth]{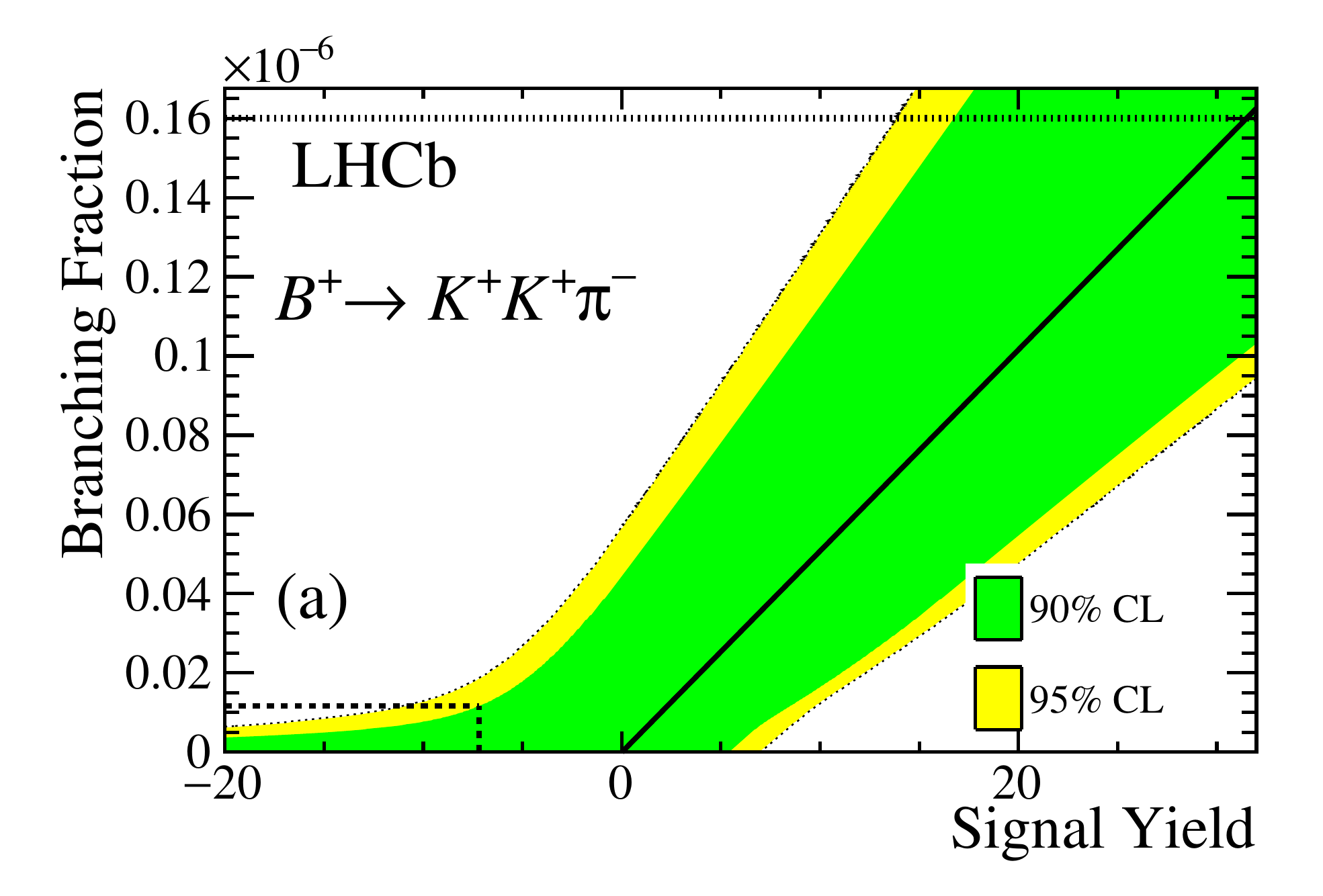} \\
  \includegraphics[width=0.78\columnwidth]{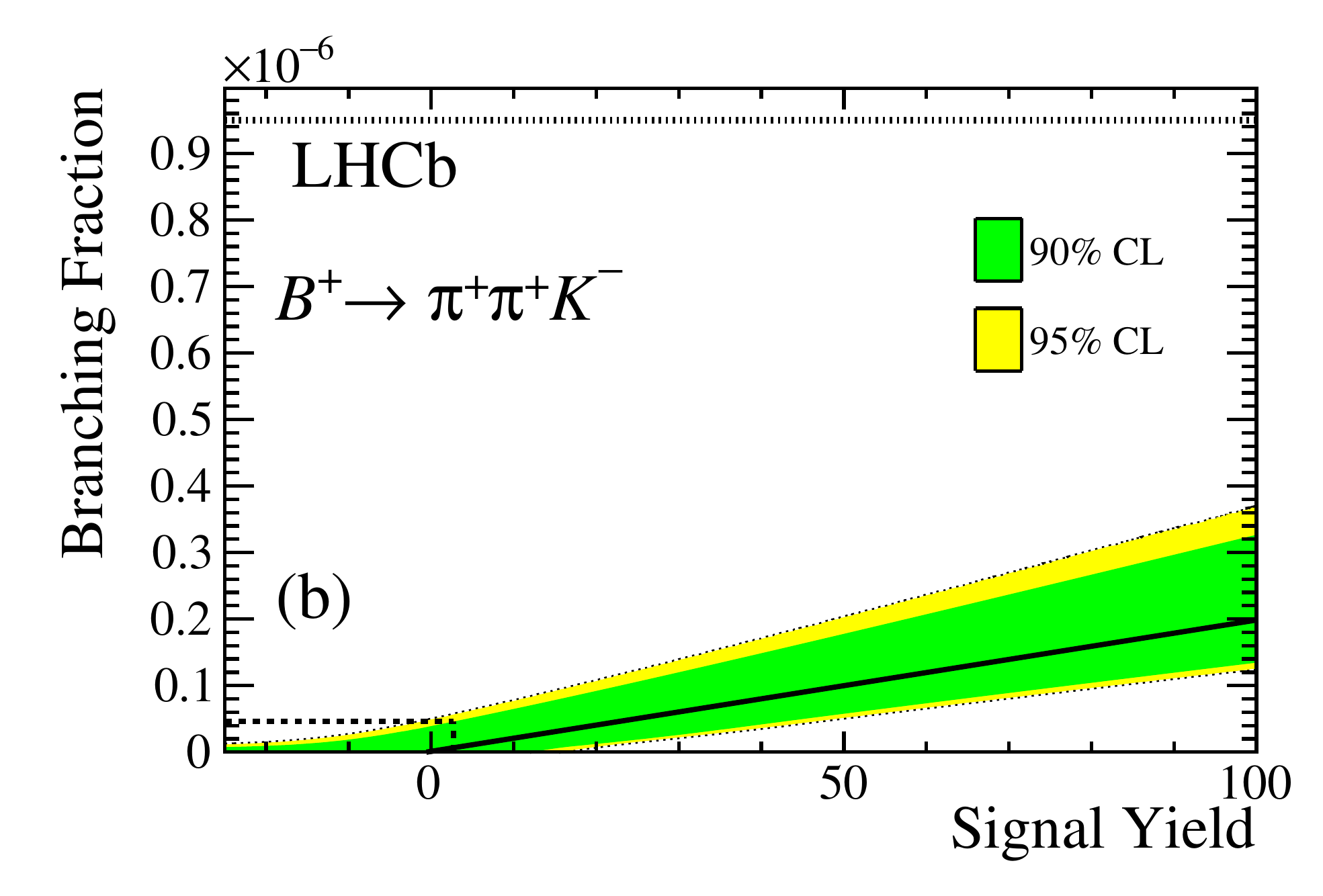}
\end{tabular}
\caption{Feldman-Cousins 90\% (green) and 95\% (yellow) confidence
  level (CL) bands for (a) \BtoKpKpPim and (b) \BtoPipPipKm, including
  statistical and systematic uncertainties. The black solid line shows
  the expected central value of the true branching fraction as a
  function of the fitted number of signal events. The horizontal
  dotted lines show the 90\% CL upper limits on the branching
  fractions prior to the present measurement. The dashed lines in the
  lower left corner of each figure show the equivalent 90\% CL upper
  limits reported in this paper.}
\label{fig:sensitivity}
\end{center}
\end{figure}

To obtain upper limits on the branching fractions,
the frequentist approach of Feldman and Cousins~\cite{Feldman:1997qc}
is used to determine 90\% and 95\% confidence region bands that relate
the true values of the branching fractions to the measured numbers of
signal events. These bands are constructed using the results of
simulation studies that account for relevant biases in the fit
procedure and include statistical and systematic uncertainties. The
construction of the confidence region bands is shown in
Fig.~\ref{fig:sensitivity}. The 90\% (95\%) confidence level (CL) upper
limits are found to be
\begin{eqnarray}
 \label{eq3}
  \BF(\BtoKpKpPim)  & < & \ULKKPiNinetyZeroFitted\ (\ULKKPiNinetyFiveFitted) \hspace{0.25cm} \mbox{at 90\% (95\%)
    CL,} \nonumber \\
  \BF(\BtoPipPipKm) & < & \ULKPiPiNinetyZeroFitted\ (\ULKPiPiNinetyFiveFitted) \hspace{0.25cm} \mbox{at 90\% (95\%)
    CL.} \nonumber
\end{eqnarray}

In summary, searches are presented for the highly-suppressed decays
\BtoKpKpPim\ and \BtoPipPipKm\ using a data sample of 3.0\invfb
collected by the \lhcb experiment in proton-proton collisions at the 
centre-of-mass energies of 7 and 8\tev. No evidence is found for these
decays and upper limits are placed on \BtoKpKpPim\ and
\BtoPipPipKm\ branching fractions. The results are approximately fourteen
and twenty times more stringent, than previous
measurements and constrain various extensions of the SM.

\section*{Acknowledgements}

\noindent We express our gratitude to our colleagues in the CERN
accelerator departments for the excellent performance of the LHC. We
thank the technical and administrative staff at the LHCb
institutes. We acknowledge support from CERN and from the national
agencies: CAPES, CNPq, FAPERJ and FINEP (Brazil); NSFC (China);
CNRS/IN2P3 (France); BMBF, DFG and MPG (Germany); INFN (Italy); FOM
and NWO (The Netherlands); MNiSW and NCN (Poland); MEN/IFA (Romania);
MinES and FASO (Russia); MinECo (Spain); SNSF and SER (Switzerland);
NASU (Ukraine); STFC (United Kingdom); NSF (USA).  We acknowledge the
computing resources that are provided by CERN, IN2P3 (France), KIT and
DESY (Germany), INFN (Italy), SURF (The Netherlands), PIC (Spain),
GridPP (United Kingdom), RRCKI and Yandex LLC (Russia), CSCS
(Switzerland), IFIN-HH (Romania), CBPF (Brazil), PL-GRID (Poland) and
OSC (USA). We are indebted to the communities behind the multiple open
source software packages on which we depend.  Individual groups or
members have received support from AvH Foundation (Germany), EPLANET,
Marie Sk\l{}odowska-Curie Actions and ERC (European Union), Conseil
G\'{e}n\'{e}ral de Haute-Savoie, Labex ENIGMASS and OCEVU, R\'{e}gion
Auvergne (France), RFBR and Yandex LLC (Russia), GVA, XuntaGal and
GENCAT (Spain), Herchel Smith Fund, The Royal Society, Royal
Commission for the Exhibition of 1851 and the Leverhulme Trust (United
Kingdom).

\addcontentsline{toc}{section}{References}
\setboolean{inbibliography}{true}
\bibliographystyle{LHCb}
\bibliography{personal,main,LHCb-PAPER,LHCb-CONF,LHCb-DP}

\newpage
\centerline{\large\bf LHCb collaboration}
\begin{flushleft}
\small
R.~Aaij$^{39}$,
B.~Adeva$^{38}$,
M.~Adinolfi$^{47}$,
Z.~Ajaltouni$^{5}$,
S.~Akar$^{6}$,
J.~Albrecht$^{10}$,
F.~Alessio$^{39}$,
M.~Alexander$^{52}$,
S.~Ali$^{42}$,
G.~Alkhazov$^{31}$
% ...

{\footnotesize \it
% ...
$ ^{5}$Clermont Universit{\'e}, Universit{\'e} Blaise Pascal, CNRS/IN2P3, LPC, Clermont-Ferrand, France\\
$ ^{6}$CPPM, Aix-Marseille Universit{\'e}, CNRS/IN2P3, Marseille, France\\
% ...
$ ^{10}$Fakult{\"a}t Physik, Technische Universit{\"a}t Dortmund, Dortmund, Germany\\
% ...
$ ^{31}$Petersburg Nuclear Physics Institute (PNPI), Gatchina, Russia\\
% ...
$ ^{38}$Universidad de Santiago de Compostela, Santiago de Compostela, Spain\\
$ ^{39}$European Organization for Nuclear Research (CERN), Geneva, Switzerland\\
% ...
$ ^{42}$Nikhef National Institute for Subatomic Physics, Amsterdam, The Netherlands\\
% ...
$ ^{47}$H.H. Wills Physics Laboratory, University of Bristol, Bristol, United Kingdom\\
% ...
$ ^{52}$School of Physics and Astronomy, University of Glasgow, Glasgow, United Kingdom\\
% ...
}
\end{flushleft}

\end{document}